\documentstyle[ltwol]{article}

\input epsf.tex  
\def\DESepsf(#1 width #2){\epsfxsize=#2 \epsfbox{#1}}

\def\be{\begin{equation}}
\def\ee{\end{equation}}
\def\bea{\begin{eqnarray}}
\def\eea{\end{eqnarray}}
\newcommand{\ltap}{\raisebox{-0.6ex}
                       {$\ \textstyle \stackrel{\textstyle <}{\sim}\ $}
                      }
\newcommand{\gtap}{\raisebox{-0.6ex}
                       {$\ \textstyle \stackrel{\textstyle >}{\sim}\ $}
                      }
\def\Bb{\textsf{\textbf{B}}}
\def\Dd{\textsf{\textbf{D}}}

\begin{document}

\title{{\it Novel} Effects in B System:
	From SUSY to Intrinsic Charm}

\author{George Wei-Shu Hou}

\address{Department of Physics, National Taiwan University,   
Taipei, Taiwan 10764, R.O.C.}   


\twocolumn[\maketitle\abstracts{ 
We have entered the era of BaBar, Belle and Tevatron competition;
with new hardware and unprecedented statistics reach,
we must be prepared for discovering new phenomena.
While these unfoldings could be coming from new physics,
it could also come about as new tricks from old.
We illustrate new physics with
generic $bsg$ dipole and its impact on $\sin2\Phi_{\phi K_S}$,
and at a deeper level,
the marriage of flavor symmetries and SUSY,
which could impact on $B_d$, $B_s$ and $D^0$ mixings and CP violation,
and possibility of a light $\widetilde{sb}$ squark.
As simple unfolding,
we touch upon charmless $B \to$ baryonic pair decay,
with or without an associated $\eta^\prime/\gamma$.
We close with the possible spectacular signal of
$B\to J/\psi D\pi$ as a flabbergasting new trick from
nonperturbative QCD: intrinsic charm of $B$.
}]

\section{New Physics Signals: Where Large?}

Traditionally,
new physics signals creep out initially as rather faint effects.
In the B Factory era (including Tevatron Run II),
we pray that new physics would emerge with a splash.
We give below three scenarios for
flavor violation in context of SUSY.

\subsection{Generic $bsg$ Dipole: \
$\sin2\Phi_{\phi K_S} \neq \sin2\Phi_{J/\psi K_S}$?}

It is known that squark-gluino loops can generate sizable 
$b_R \to s_L g$ transitions,
which probes a possible new CP phase \cite{HT} 
associated with $b_R$ that is not probed by $V_{\rm CKM}$.
Parametrizing the dipole strength as 
$c_{11} = \vert c_{11}\vert e^{i\sigma}$,
the coupling was employed to
enhance the direct CP asymmetries ($a_{\rm CP}$) 
in $B^0\to K^+\pi^-$ mode, rumored to be sizable in late 1997.
By interfering destructively with SM penguins to satisfy
$B^+\to \phi K^+ \ltap 5\times 10^{-6}$ from CLEO, \cite{phiK}
it was found \cite{HHY98} that $a_{\rm CP}(K^+\pi^-) > 50\%$ is possible.
Subsequently, CLEO reported \cite{aCP}
no evidence for $a_{\rm CP}(K^+\pi^-)$.
This diminishes, but does not eliminate, the
prospects for direct CP in $\phi K$ mode,
especially since Belle discovered       \cite{bozek} that
$B^+\to \phi K^+$ is considerably above the old CLEO bound.
Besides $a_{\rm CP}$,
we are now more interested in 
mixing-dep. CP in $\phi K_S$ mode.
Taking as illustration
that $b\to sg \sim 2.5\%$,
which is 10 times larger than SM but very hard to rule out,
we find \cite{HY} that $\Phi_{\phi K_S}$ could be shifted by $\sim
20^\circ$, leading to e.g.  $\sin2\Phi_{\phi K_S} \simeq 0.93$ for 
$\sin2\Phi_{J/\psi K_S} \simeq 0.48$ (the Belle value). \cite{Belle}

\subsection{Generic Abelian Flavor Symmetry with SUSY}

New physics in {\it flavor} sector is likely
since little is understood.
The intriguing pattern of mass and mixing hierarchies
in powers of $\lambda \equiv \vert V_{us}\vert$ suggest
\be
{M_u\over m_t} \sim \left[
\matrix{\lambda^7 &\lambda^5 &\lambda^3 \cr
         ? &\lambda^4 &\lambda^2 \cr 
         ? & ? &1}
\right],\ \ \
{M_d\over m_b}\sim % \lambda^3 
\left[
\matrix{\lambda^4 &\lambda^3 &\lambda^3 \cr
         ? &\lambda^2 &\lambda^2 \cr 
         ? & ? &1}
\right],
\ee
where the upper right is from
$U_L$, $D_L \sim V_{\rm CKM} \equiv U_L^\dagger D_L$ 
which holds in suitable basis.
Note that the lower left are diagonalized by $U_R$, $D_R$
but unknown to us with SM dynamics only.
Eq. (1) clearly suggest some 
possible underlying flavor (horizontal) symmetry. 
If this symmetry is Abelian,
commuting horizontal charges imply
$M_{ij} M_{ji} \sim M_{ii} M_{jj}$
($i$, $j$ not summed),
hence
\be
{M_u\over m_t} \sim \left[
\matrix{\lambda^7 &\lambda^5 &\lambda^3 \cr
         \lambda^6 &\lambda^4 &\lambda^2 \cr 
         \lambda^4 &\lambda^2 &1}
\right],\ \ \ 
{M_d\over m_b}\sim % \lambda^3 
\left[
\matrix{\lambda^4 &\lambda^3 &\lambda^3 \cr
         \lambda^3 &\lambda^2 &\lambda^2 \cr 
         \lambda &1 &1}
\right],
\ee
is inferred.
It is intriguing, then,
that
{\it $M_d^{32}/m_b$, $M_d^{31}/m_b$ are
the most prominent off-diagonal elements},
hence impact on $B_d$ and $B_s$ mixings naturally,
 {\it iff} right-handed down sector can be heard.
However, the SM has no right-handed flavor dynamics.
This is where SUSY enters to help: $\tilde d_R$ couples to $\tilde g$.

Assuming that SUSY breaking itself does not introduce
flavor violations,
we find that
$(\widetilde M^2_q)^{ij}_{LR}=(\widetilde M^2_q)_{RL}^\dagger
\sim\widetilde m\,M_d^{ij}$,
${(\widetilde M^2_Q)_{LL}} \sim~\widetilde m^2 V_{CKM}$, but
\be
(\widetilde M^{2}_d)_{RR}  
\sim \widetilde m^2 \left[  
\matrix{1 &\lambda &\lambda \cr  
         \lambda &1 &1 \cr   
         \lambda &1 &1}  
\right],      
\ee
contribute significantly to $B_d$ (or $B_s$) mixings.

Generic flavor symmetry and its breaking
can impact on measurables via SUSY!
We stress that the flavor and CP violation in Eq. (3)
are on the same footing as $V_{\rm CKM}$.
%We now explore the two cases of $d_R$-$b_R$ and $s_R$-$b_R$ mixings.

%
\subsubsection{$d_R$-$b_R$ Mixing: \  
 Low $\sin2\Phi_{B_d}$ and $D^0$ Mixing?}

The $RR$ sector could   
contribute significantly to $B_d$ mixing
via $\delta^{13}_{dRR}\sim\lambda$
since this is much larger than
$V_{td} \sim \lambda^3$.
A simple dimensional analysis
suggests that $\widetilde m$,
 $m_{\tilde g} \sim M_W/\lambda^2 \sim$ TeV scale could
generate squark-gluino box diagram contributions
that are comparable to SM.
We illustrate \cite{CH} this observation in Fig. 1,
where
$\sin2\Phi_{B_d}$ via $J/\psi K_S$ 
can range from 0.3 to 1
vs $\sin2\phi_1 \simeq 0.75$--$0.71$ 
for $\phi_3 = 65^\circ$--$85^\circ$ in SM.

%%%%%%%
\begin{figure}[htb]  
\centerline{{\epsfxsize1.65 in \epsffile{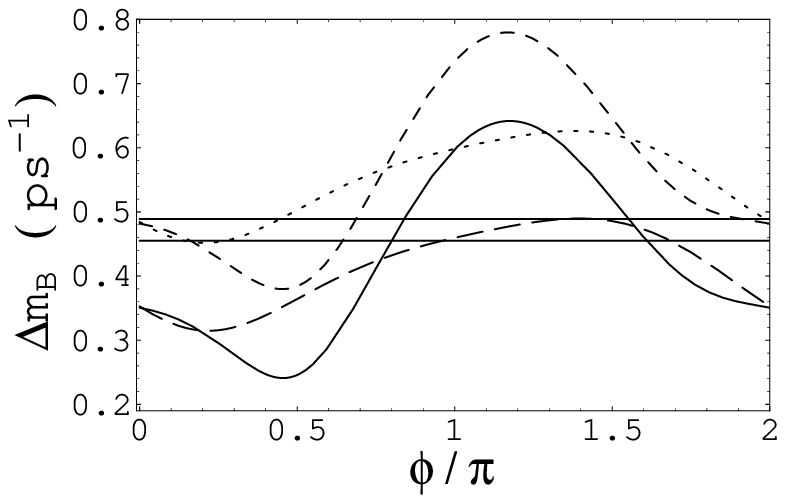}}  
            {\epsfxsize1.65 in \epsffile{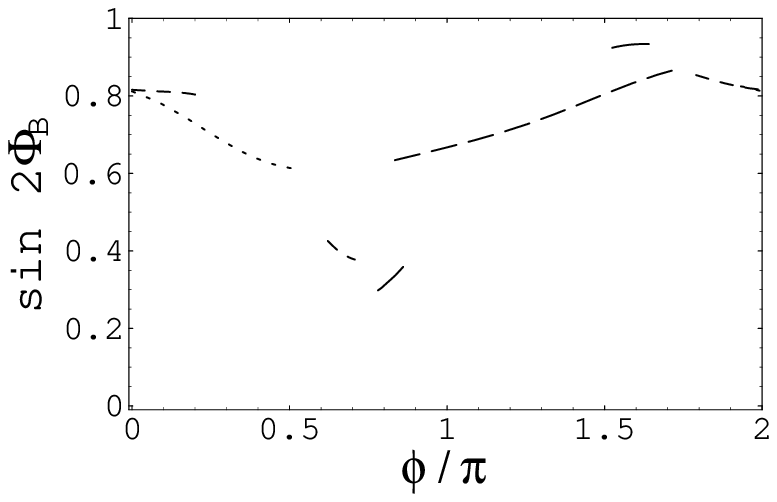}}}  
\vskip-0.2cm%\smallskip  
\caption{
$\Delta m_{B_d}$ and $\sin2\Phi_{B_d}$
 vs. $\phi \equiv \arg \delta_{dRR}^{13}$, 
including both SM and SUSY effects, 
for squark mass $\widetilde m =$ 1.5 TeV
(and $\tan\beta =2$ and $|\mu|<m_{\tilde g}$). 
Horizontal double lines indicate $2\sigma$ experimental range.
Solid (short-dash), long-dash (dotted) curves 
for $m_{\tilde g} = 1.5$, 3 TeV 
and $\phi_3 = 65^\circ$ ($85^\circ$), respectively.
}\vskip-0.35cm
\label{fig:db}  
\end{figure}

Of particular interest is
the low $\sin2\Phi_{B_d} \sim 0.3$--$0.4$ possibility,
stated already   \cite{CH} in May 2000 (before ICHEP2000),
as compared with the present
world average of $0.48\pm 0.16$,
dominated by BaBar ($0.34\pm 0.20\pm 0.05$) and
 Belle ($0.58^{+0.32+0.09}_{-0.34-0.10}$) values
reported at this conference.
It is clear that CKM unitarity bound from
$\Delta m_{B_s}/\Delta m_{B_d}$ should be relaxed,
and potential conflict on $\phi_3/\gamma$ w.r.t.
charmless rare $B$ decays may be alleviated.
What we mean is that,
with $\widetilde m$, $m_{\tilde g} \gtap$ TeV
and $(\widetilde M^2_q)_{LR,RL}$ 
suppressed by $m_q/\tilde m$,
there is little impact on penguins,
hence charmless rare $B$ decays may have better access to CKM phases
(except for hadronic uncertainty).
Thus,
$\phi_3/\gamma \gtap 90^\circ$ may well be the case, \cite{HHY}
which is strengthened by 
$\pi^+\pi^-/K^+\pi^- \sim 1/4$ as reported
by CLEO, Belle and now BaBar at this conference.

We eagerly await summer results on $\sin2\Phi_{B_d}$!!

But we have been too naive so far: 
$\Delta m_K$ and $\varepsilon_K$ constraints are much more stringent.
It is impossible to sustain $\delta^{12}_{dLL,RR} \sim \lambda$,
even with $\widetilde m$, $m_{\tilde g} \gtap$ TeV.
Traditionally one employs quark-squark alignment (QSA) to 
impose ``{\it texture zeros}" on quark mass matrices,
i.e. $M_d^{12,21}=0$ 
hence $D^{12}_{L,R}=0$ or highly suppressed.
%In this way,
%$\delta^{12}_{dLL,RR}$ can be suppressed
%hence kaon mixing constraint satisfied.

%%%%%%%
\begin{figure}[b]
\vskip-0.3cm
\centerline{{\epsfxsize1.8 in \epsffile{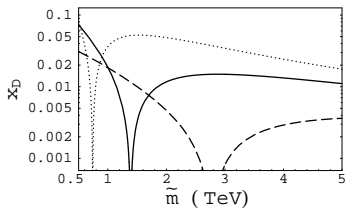}}
}
\vskip-0.2cm%\smallskip  
\caption{
$x_D$ vs $\tilde m$ for 
$m_{\tilde g}= 0.8$, 1.5 and 3 TeV
and $\tan\beta = 2$.
}
\label{fig:uc}  
\end{figure}

In so doing, however,
one notices that $D^{12}_{L} \simeq 0$ implies
$U^{12}_{L}\sim \vert V_{cd}\vert =\lambda$,
which is a general consequence of QSA.
Thus, $\tilde u_L$-$\tilde c_L$ mixing 
$\delta^{12}_{uLL} \sim \lambda$ is sizable,
which can generate $D^0$--$\bar D^0$ mixing,
{\it right in the ballpark of recent tantalizing hints} 
from the CLEO and FOCUS experiments, $x_D \sim 0.01$.
Note that the zeros in Fig. 2 reflect cancellation when 
different terms have common phase,
and shows that $x_D$ can be considerably below 0.01.
In any case it is exciting that
$D^0$ mixing at such levels can be further studied at Belle and BaBar.

There is an interesting subtlety for 
our choice of $M_d^{31}\neq 0$
if one wishes to retain $M_d^{23,32}$:
 $M_d^{12,21}$ would once again be generated.
Thus, if we choose to keep 
$(\widetilde M^2_d)_{RR}^{13}/\widetilde m^2 \sim \lambda$
then $M_d^{23} = M_d^{32} =0$ need to be imposed
on top of $M_d^{12} = M_d^{21} =0$
and the $s$ flavor is decoupled from $d$, $b$,
hence there will be no new physics effects
in $B_s$ mixing and $b\to s\gamma$ decays!
We seem to find that
the stringent $\Delta m_K$ and $\varepsilon_K$ constraints
imply 4 texture zeros in $M_d$.
We now turn briefly to the case of
decoupling $d$ flavor with QSA.
For a more generic discussion of SUSY violation impact on
$\sin2\Phi_{B_d}$, $B\to \pi\pi$ and $\rho\gamma$,
see the poster talk of C.K. Chua. \cite{Chua}

\subsubsection{$s$-$b$ Mixing: \ 
 $\Delta m_{B_s}$, $\sin2\Phi_{B_s}$; Light $\widetilde{sb}$ Squark}

I will be brief since this subject is 
covered by the poster talk of A. Arhrib. \cite{Arhrib}
The previous $d$-$b$ mixing case satisfy $\Delta m_K$, $\varepsilon_K$
 by construction (via alignment),
but still have interesting, measurable effects in
$B_d$ and $D^0$ mixings, even if SUSY particles are at TeV scale.
The reason is
the large $\tilde d_R$-$\tilde b_R$
and $\tilde u_L$-$\tilde c_L$ mixings ($\sim \lambda$)
that arise from Abelian horizontal charges and low energy constraints.
Unfortunately,
the SUSY scale becomes so high, practically 
there can be no impact on penguins,
hence $\varepsilon^\prime/\varepsilon$, $b\to s\gamma$ and $b\to d\gamma$
are all unaffected.
Though viable, the case is depressing in that
squarks and gluino cannot be produced at Tevatron or even the LHC,
while there is also no impact on $B_s$ System!

Changing the mindset, however, one could have
interesting phenomena in a rather similar context:
$s$-$b$ mixing ($\heartsuit$)!
Decoupling $d$ flavor now with QSA,
one finds,
\be
(\widetilde M^{2}_d)_{RR}  
\sim \widetilde m^2 \left[  
\matrix{1 &0 &0 \cr  
         0 &1 &1 \cr   
         0 &1 &1}  
\right],
\ee
where one has 4 texture zeros analgous to $d$-$b$ case,
but {\it $\tilde s_R$-$\tilde b_R$ mixing is $\sim 1$}!
The exciting new feature from the see-saw pattern of Eq. (4) is that, 
one down type squark, which we call 
the {\it strange-beauty squark} $\widetilde{sb}_1$, 
could be {\it driven light by the large $s$-$b$ mixing}! 
This is rather different from the scenarios of 
light stop or sbottom which are generated by
large top Yukawa coupling (with or without large $\tan\beta$),
and represents a third mechanism for having 
one squark much lighter than the rest,
in this case as arising from flavor violation in right-hand sector.

It is truly intriguing that,
even for $m_{\widetilde{sb}_1}$ and neutralino 
(dominantly bino) mass $m_{\widetilde \chi_1^0}$ as light as 100 GeV,
penguins are still little affected,
and $b\to s\gamma$ is quite accommodating \cite{Arhrib}
even with large $\tilde s_R$-$\tilde b_R$ mixing!
But the impact on $B_s$ mixing and $\sin2\Phi_{B_s}$,
whether $\widetilde{sb}_1$ is light or not,
is rather visible,
as one can easily see by scaling up from $B_d$ result
for $d$-$b$ mixing case.
Thus, once again the $\Delta m_{B_d}/\Delta m_{B_s}$ constraint
should be loosened, this time due to $\Delta m_{B_s}$ being affected.
We note that $D^0$ mixing remains volatile and interesting
because of alignment.

If the $\widetilde{sb}_1$ is in fact light
and with $\widetilde \chi_1^0$ as LSP,
there is a change in signature for collider search.
Since $\widetilde{sb}_1$ has roughly equal mixture of
$s$ and $b$ flavor,
one should keep in mind that
$\widetilde{sb}_1 \to b\tilde\chi_1^0$, $s\tilde\chi_1^0$
are both present, hence $b$-tagging is less efficient.
Thus, the direct bound on $\widetilde{sb}_1$
should be weaker than the standard $\tilde b$ squark.

We find with interest that the signatures of
$\Delta m_{B_s}$, $\sin2\Phi_{B_s}$ and direct $\widetilde{sb}_1$ search
can all be conducted at the Tevatron Run II,
while $D^0$ mixing as well as the CKM phase angle pattern
can be studied at the B Factories.

\section{Rare Baryons: \ New Pathways?}

Charmless rare {\it mesonic} modes
started to emerge in 1997,
with many modes now with measured rates $> 10^{-5}$.
Charmless rare {\it baryonic} modes are far less fruitful.
We have only the CLEO98 bounds of
$B\to \bar\Lambda p$, $\bar\Lambda p\pi^-$, $\bar pp
 < 0.26$, 1.3, 0.7 $\times 10^{-5}$ based on 5.8M $B\bar B$'s.
The corresponding theory is equally sparse:
just a handful of models
that were ``stimulated" by the old ARGUS false observation of
$B\to p\bar p(\pi)$ in the late 1980's.

{\it Where is the best place to search?}

%\subsection{New Tricks: \
%$B\to \eta^\prime{\Bb}_s\bar{\Bb}$, $\gamma{\Bb}_s\bar{\Bb}$}

Observation: Smallness of $B\to \bar\Bb_{(s)}\Bb$ 
likely rooted in the {\it large energy release},
aggravated by {\it more complicated composition} of 
baryons ($qqq$) vs mesons ($q\bar q$).
In particular,
the 4-quark operators that mediate $b$ decay
quite naturally project a $B$ meson onto a pair of $q\bar q$ quarks
in final state.
Thus, to find larger charmless baryonic $B$ decays, one needs \cite{HS}
 1) {\it reduced energy release} and
 2)~{\it baryonic ingredients in final state}.

From these insights,
we suggest the natural starting points as:
Inclusive $B\to \eta^\prime + X_s$ and $\gamma + X_s$.
Both cases start with large rates,
the former $\sim 6\times 10^{-4}$ for $p_{\eta^\prime} > 2.0$ GeV,
while the latter $\sim 2\times 10^{-4}$ for $p_{\gamma} \gtap 2.0$ GeV.
Both processes have $\eta^\prime/\gamma$ carry away large energy,
hence reduced energy release in the recoil $X_s$ system!

%%%%%%%
\begin{figure}[b]
\vskip-0.3cm
\centerline{{\epsfxsize2.5in \epsffile{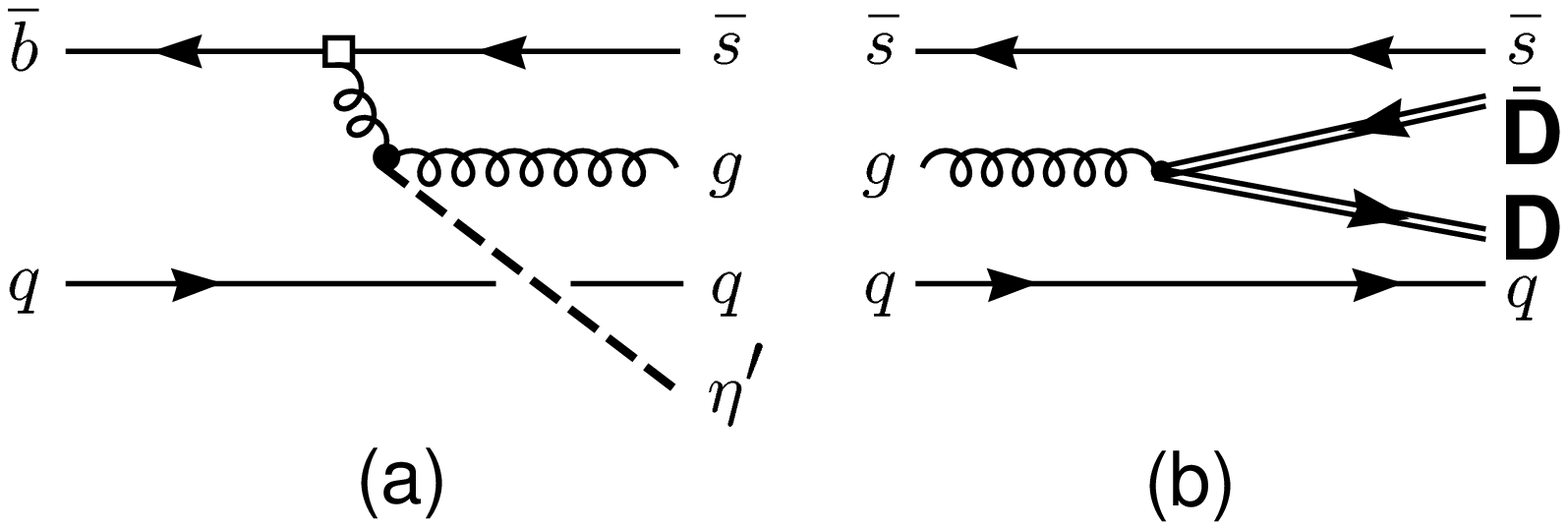}}
}
\vskip-0.3cm
%\smallskip  
\caption{
Anomaly motivated two step process:
(a) $g^* \to g^*\eta^\prime$ mechanism for fast $\eta^\prime$ emission;
(b) $g^* \to \bar \Dd \Dd$ diquark pair.
}
\label{fig:bsetapDDbar}  
\end{figure}

From an inclusive picture of charmless baryon formation,
we envision the anomaly mechanism       \cite{HT} which is effective at 
spitting out energetic $\eta^\prime$ mesons (Fig. 3(a)),
followed by $g^*\to \bar \Dd \Dd$ splitting of gluon into diquark pair.
In this way, as can be seen from Fig. 3(b),
we have baryonic pair ingredients in final state.
We then allow a phase space argument for baryon pair formation (Fig. 4).

%%%%%%%
\begin{figure}[tb]
\centerline{{\epsfxsize2.5in \epsffile{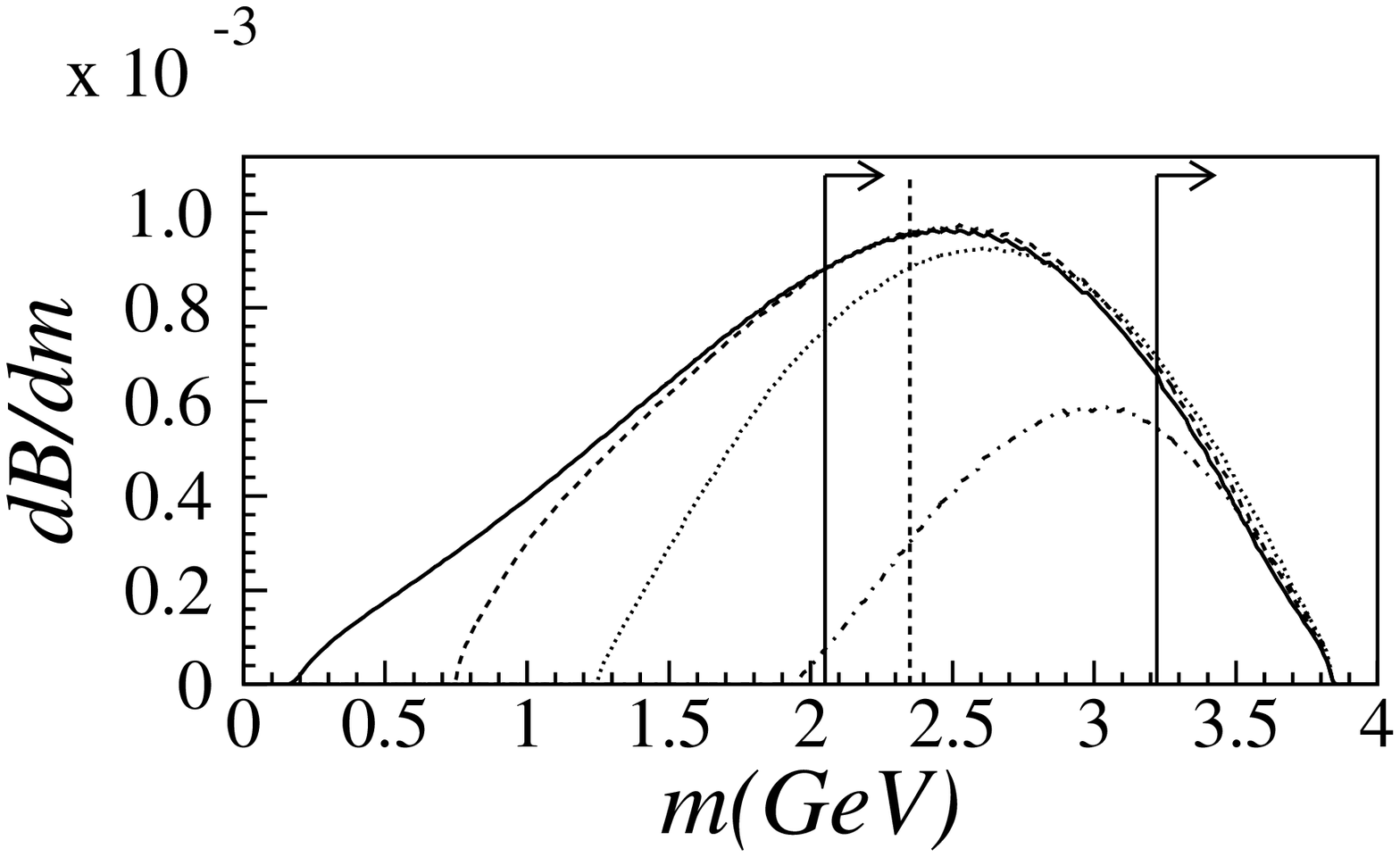}}
}
\vskip-0.3cm
%\smallskip  
\caption{
$\bar B\to  \eta^\prime + s g \bar q
\to \eta^\prime\Bb_{(s)}\bar\Bb(\pi)$ from Fig. 3
with solid, dash, dots, dotdash 
$\sim m_g = 0,\ 0.6,\ 1.1,\ 1.8$ GeV in phase space.
}\vskip-0.35cm
\label{fig:BetapBbBbbar}  
\end{figure}

Since $\bar \Dd \Dd$ pairs already appear to left of
$m_g \sim 1.1$ GeV (dots) in Fig. 4,
while $\bar \Lambda N$ threshold opens up only at 2.05~GeV
(left vertical line with arrow),
we expect threshold enhancement for $s g \bar q \to
\Bb_{(s)}\bar\Bb(\pi)$ around $m_{X_s} \sim $ 2.3~GeV,
which corresponds to the experimental cut on 
$K+n\pi$ partial reconstruction.
The modes to search for are $\bar B\to \eta^\prime \Lambda \bar N$
and similar low lying $\bar\Bb_{s}\Bb$ states,
together with relatively fast $\eta^\prime$.
Since reconstruction is easy and background
is expected to be low ($\Lambda_c^+\bar N$ threshold at 3.22 GeV),
the process may offer important 
probe into higher mass $m_{X_s}$ spectrum
(the envelope that drops beyond $m_{X_s}$ beyond 2.5 GeV)
that is important for confirming the anomaly mechanism itself.

%%%%%%%
\begin{figure}[b]
\vskip-0.5cm
\centerline{{\epsfxsize2.5in \epsffile{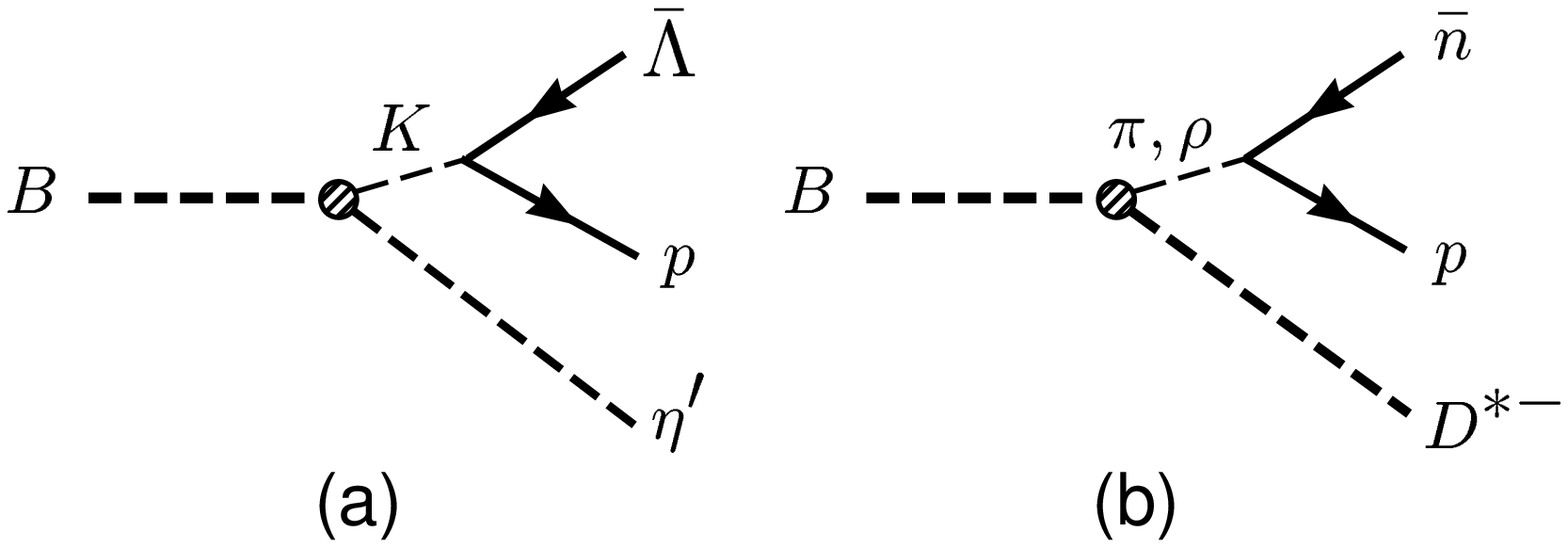}}
}
\vskip-0.3cm
%\smallskip  
\caption{
Analogy between
(a) $B\to \eta^\prime p\bar\Lambda$ and (b) $B\to D^{*-} p\bar n$
}
\label{fig:Dstarpnbar}  
\end{figure}

Further encouragement is obtained by improving \cite{CHT}
the pole model approach \cite{HS} by making analogy (see Fig. 5) of 
$B\to \eta^\prime p\bar\Lambda$ with the recently reported 
$B\to D^{*-} p\bar n$ mode by CLEO. 
Assuming factorization, using $B\to D^*$ form factors
and incorporating proton form factor (FF),
the vector current part can account for $\sim $ half the
observed rate,
with the other half 
presumably through axial-vector (e.g. $a_1$) channel.
Extending to $B \to\eta^\prime\bar\Lambda p$,
 $\gamma\bar\Lambda p$, perhaps even $\ell\nu\bar NN$,
we caution that there is no analogy to proton FF,
but this may actually imply a larger effect.
We therefore suggest \cite{HS} that 
$B\to \eta^\prime\bar\Lambda p$,
$\gamma\bar\Lambda p \sim 10^{-5} > \bar \Lambda p$ as plausible,
and may be the {\it first charmless baryon mode(s)} 
to be observed.
One has the extra bonus of self-analyized spin in 
$\Lambda \to p\pi$ decay,
which may probe $B\to \eta^\prime,\; \gamma$ dynamics
via the CP odd and even
$\Delta_{\rm odd,\ even} = \kappa_{\bar \Lambda}
			   \mp \kappa_{\Lambda}$,
where $\kappa_{\Lambda} = ${\boldmath $s$}$_{\Lambda}\cdot$
({\boldmath $p$}$_p \times${\boldmath $p$}$_{\Lambda}$
and $\kappa_{\bar \Lambda} = ${\boldmath $s$}$_{\bar \Lambda}\cdot$
({\boldmath $p$}$_p \times${\boldmath $p$}$_{\bar \Lambda})$
are both T-odd.
New physics may be eventually uncovered by such triple products.

Of course, search for traditional $B\to \bar \Bb_{(s)}\Bb$ 2-body modes
should continue!
Unlike $K\pi > \pi\pi$,
we find that $\bar{B^0} \rightarrow \Sigma^{+} \bar{p}
 \ltap \bar{B^0} \rightarrow p \bar{p}$
as the two leading modes.

%%%%%%%
\begin{figure}[t]
\centerline{\epsfxsize1.5in \hskip0.4cm \epsfbox{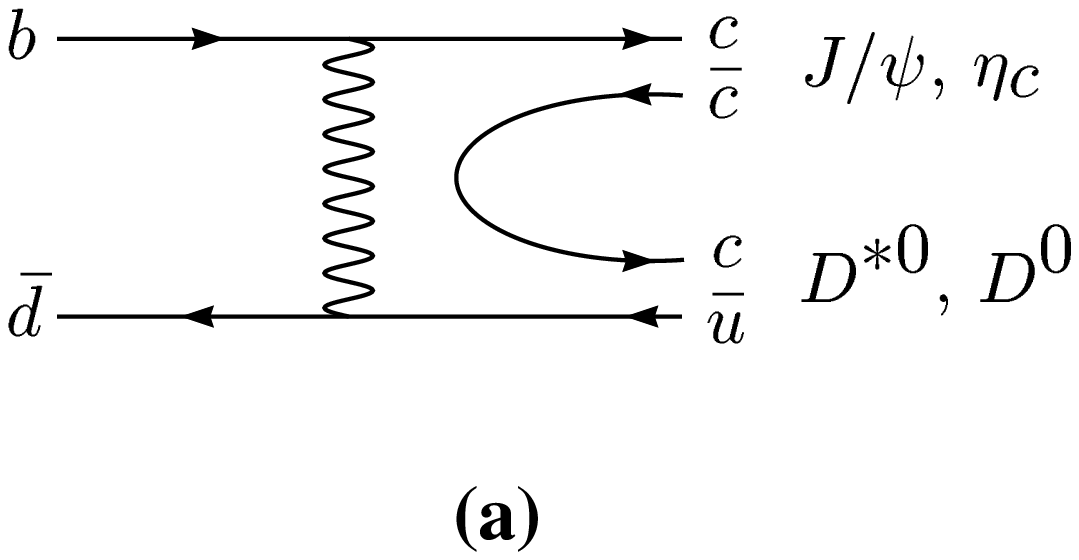} \hskip0.2cm
            \epsfxsize1.5in \epsfbox{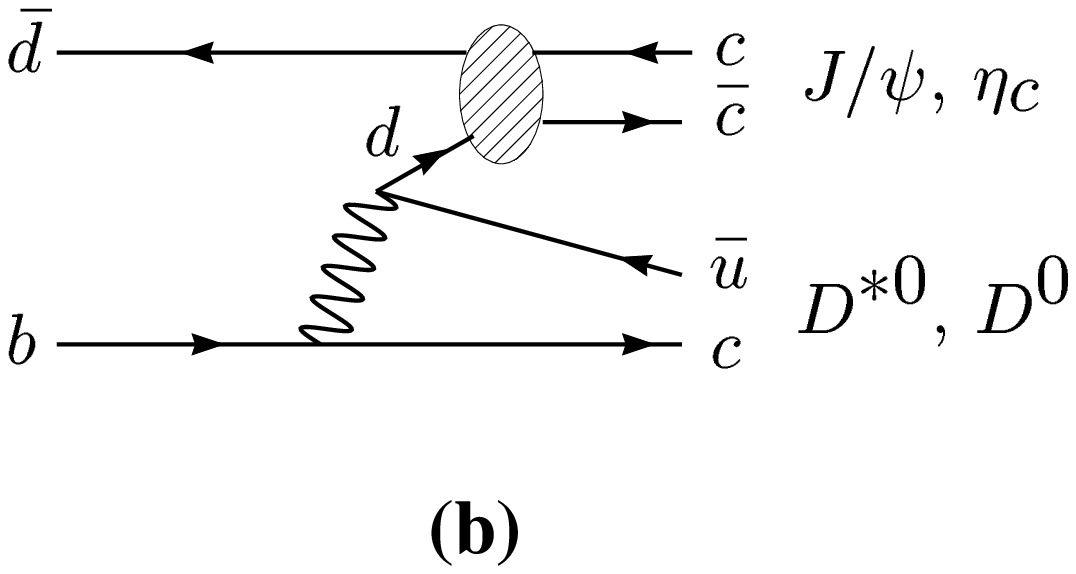}}
%\vskip0.1cm
\centerline{\epsfxsize1.5in \hskip0.4cm \epsfbox{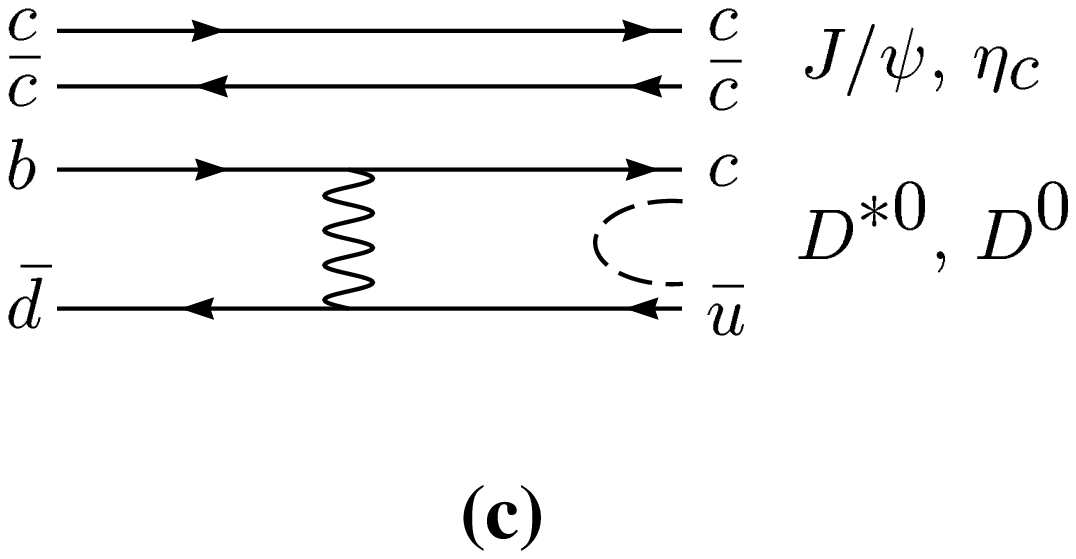} \hskip0.2cm
            \epsfxsize1.5in \epsfbox{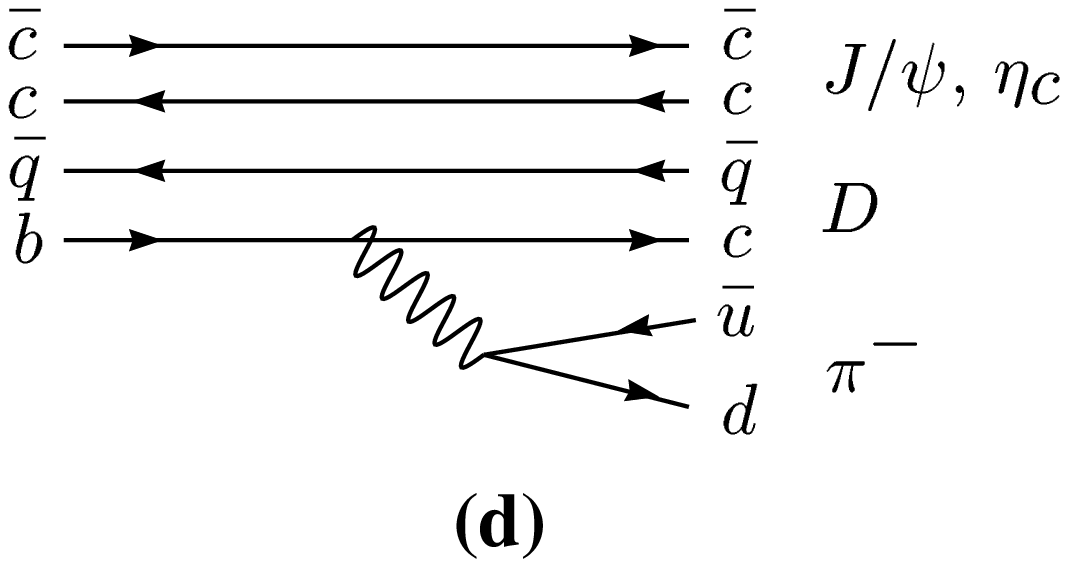}}
%\vskip0.1cm
\centerline{\epsfxsize1.3in \epsfbox{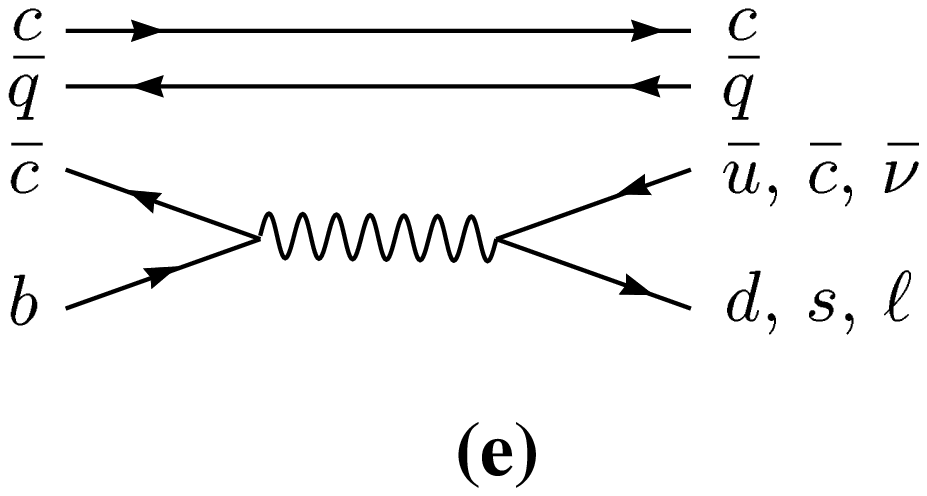} \hskip0.7cm
            \epsfxsize1.3in \epsfbox{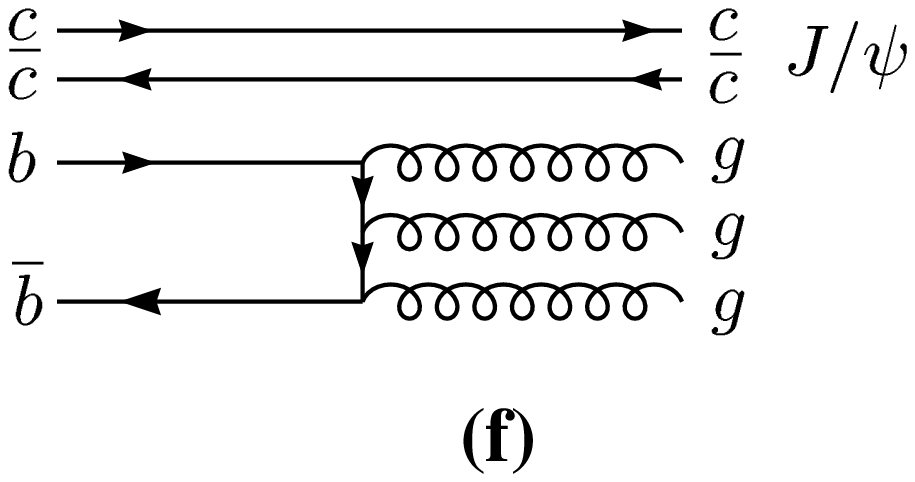}}
\vskip-0.2cm
\caption{
(a), (b) Suppressed $B\to J/\psi D$ decay;
(c)-(f) $B$, $B_c$ and $\Upsilon$ decay 
via intrinsice charm Fock component.
}\vskip-0.35cm
\label{fig:IC_Feyn}  
\end{figure}

\section{Intrinsic Charm: $B\to J/\psi \, D\pi$!?}

At first sight, this seems like a red herring:
3 charm quarks in $B$ decay final state!
Clearly, the exchange and OZI violating processes
of Figs. 6(a) and (b) should be extremely suppressed.
This could be an advantage, however,
depending on what intrinsic charm (IC) could do. \cite{IC}
Naively, Fig. 6(c) could lead to
$\bar B^0\to J/\psi D^{(*)0}$, but would probably suffer from 
annihilation suppression, while for Fig. 6(d),
normal spectator picture leads to
$\bar B\to J/\psi D \pi^-$, but is close to the phase space limit.
The question is the distribution and strength of IC.

The distributions can be readily obtained, as is shown in Fig. 7.
The strength of IC is of nonperturbative origin,
and cannot be deduced from first principles yet.
An analysis of EMC data indicates that 
IC of proton could be $\sim 0.86\%$.
Since the $B$ meson is smaller than
the proton because of a larger reduced mass,
we expect a larger IC fraction in $B$ than in light hadrons!
We may therefore hope for IC in $B$ to be greater than 1\%.

%%%%%%%
\begin{figure}[b]
\vskip-0.6cm
\centerline{ \hskip0.2cm{\epsfxsize1.5in \epsfbox{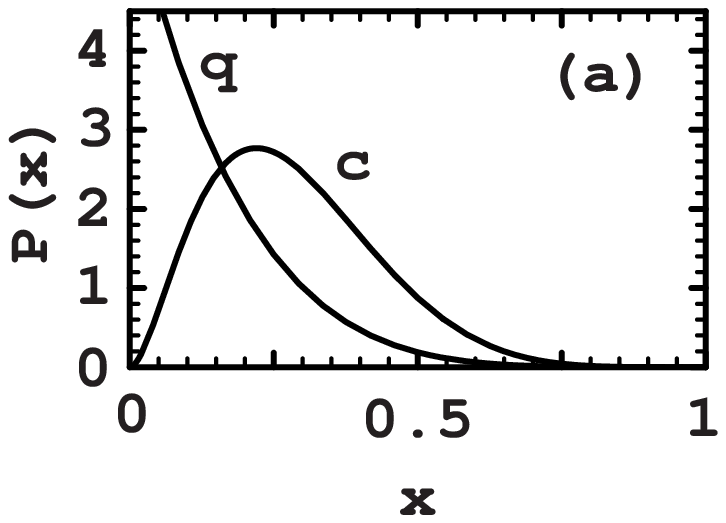} \hskip-0.3cm
            \epsfxsize 1.5in \epsfbox{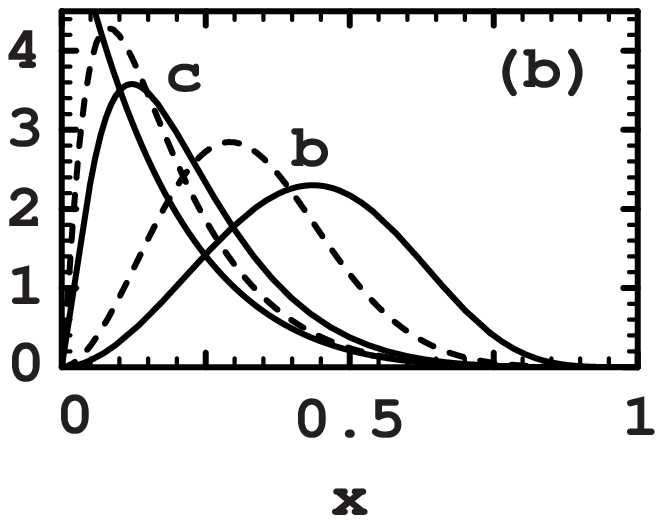}}
}
\vskip-0.3cm
%\smallskip  
\caption{
IC in 
(a) $p$ and (b) $B$ meson (dashes for $\Upsilon(1S)$).
}
\label{fig:ICdist}  
\end{figure}

%%%%%%%
\begin{figure}[t]
\vskip-0.15cm
\centerline{ {\epsfxsize2.1in \epsfbox{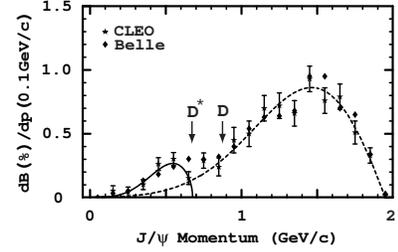} }
}
\vskip-0.2cm
%\smallskip  
\caption{
Feed-down subtracted inclusive $B\to J/\psi +X$.
}\vskip-0.3cm
\label{fig:slowJpsi}  
\end{figure}

It is exciting \cite{IC} that
the above scenario may already have some experimental bearing.
Published CLEO data as well as new Belle results from ICHEP2000
hint at an excess of low momentum $J/\psi$ events from $B$ decay.
The curves in Fig. 8 are simple fits,
with solid for excess below 0.9 GeV assuming
$D\pi$ recoil starting at 0.66~GeV, the $D^*$ recoil threshold.
We see that $\bar B\to J/\psi D \pi^-$
from Fig. 6(d) provides a plausible explanation for this excess.
There may be a hint of $\bar B\to J/\psi D^*$,
but there is no indication for $J/\psi D$.
The plausibility is enhanced when we find that
an IC at 1\% level or higher,
with distribution as indicated in Fig. 7, 
can account for \cite{IC} the rate of few $\times 10^{-4}$.
The search should be straightforward,
and verification could be as early as this summer.

We note that the process of Fig. 6(f) may explain the soft spectrum
of $\Upsilon(1S) \to J/\psi +X$,
where $p_{J/\psi}$ peak at $\sim 1.5$ GeV.
It would be amusing if
smoking gun evidence for IC emerges at B Factories,
rather than for lighter hadrons.

\vskip 0.3cm  
\noindent{\bf Acknowledgement}.\ \  
I have enjoyed collaborating with
Abdes Arhrib,
Chia-Hung Chang,
Chun-Khiang Chua, 
Amarjit~Soni, 
Shang-Yuu Tsai and
Kwei-Chou Yang,
as well as earlier collaborators.

\end{document}